\documentclass[conference]{IEEEtran4PSCC}

\IEEEoverridecommandlockouts
\usepackage{cite}
\usepackage{amsmath,amssymb,amsfonts}
\usepackage{algorithmic}
\usepackage{graphicx}
\usepackage{textcomp}
\usepackage[caption=false,font=footnotesize]{subfig}
\usepackage{color}
\usepackage{xcolor}
\usepackage{cite}
\usepackage{amsmath}
\usepackage{amssymb}
\usepackage{amsthm}
\usepackage{graphicx}

\usepackage{epstopdf}
\usepackage{comment}
\usepackage{multirow}
\usepackage{paralist}
\usepackage{lineno}
\usepackage{mdwlist}
\usepackage{eurosym}\DeclareGraphicsExtensions{.pdf,.png,.jpg}
\usepackage{breqn}
\usepackage{makecell}
\usepackage{soul}
\usepackage{empheq}
\graphicspath{{./pic/}}
\usepackage[super]{nth}
\usepackage{subfig}
\usepackage{bm}
\usepackage[normalem]{ulem}
\usepackage{xcolor}
\def\BibTeX{{\rm B\kern-.05em{\sc i\kern-.025em b}\kern-.08em
    T\kern-.1667em\lower.7ex\hbox{E}\kern-.125emX}}

\newcommand\upj{\mathord{\mathrm{j}}}
\usepackage{accents}

\begin{document}
\title{Topology-aware Piecewise Linearization of the AC Power Flow through Generative Modeling}
\author{\IEEEauthorblockN{Young-ho Cho and Hao Zhu}
\IEEEauthorblockA{Chandra Department of Electrical and Computer Engineering \\
The University of Texas at Austin\\
Austin, TX, USA \\
\{jacobcho, haozhu\}@utexas.edu}
}
\maketitle

\begin{abstract}
    Effective power flow modeling critically affects  the ability to efficiently solve large-scale grid optimization problems, especially those with topology-related decision variables.
    In this work, we put forth a generative modeling approach to obtain a piecewise linear (PWL) approximation of AC power flow by training a simple neural network model from actual data samples. By using the ReLU activation, the NN models can produce a PWL mapping from the input voltage magnitudes and angles to the output power flow and injection. Our proposed generative PWL model uniquely accounts for the nonlinear and topology-related couplings of power flow models, and thus it can greatly improve the accuracy and consistency of output power variables. Most importantly, it enables to reformulate the nonlinear power flow and line status-related constraints into mixed-integer linear ones, such that one can efficiently solve grid topology optimization tasks like the AC optimal transmission switching (OTS) problem.  Numerical tests using the IEEE 14- and 118-bus test systems have demonstrated the modeling accuracy of the proposed PWL approximation using a generative approach, as well as its ability in enabling competitive  OTS solutions at very low computation order. 
\end{abstract}

	
\thanksto{\protect\rule{0pt}{0mm}
This work has been supported by NSF Grants 1802319 and 2130706.}

\section{Introduction}

\IEEEPARstart{E}{ffective} power flow modeling is critical  for analyzing and optimizing large-scale power systems for efficient and reliable grid operations. With worldwide energy transitions and decarbonization, grid optimization tasks  are increasingly  challenged by e.g., uncertainty factors, extreme conditions, and fast computation needs. Meanwhile, optimizing the grid topology for increased flexibility  has been advocated in problems like optimal transmission switching (OTS) \cite{fisher2008optimal}, adaptive islanding \cite{trodden2013optimization}, and post-disaster restoration  \cite{chen2017sequential}. Hence, it is important to develop effective power flow models that can facilitate the accurate and fast solutions for grid optimization problems, especially those with the combinatorial topology variables.   


There exist significant efforts in developing approximation models for the nonlinear AC power flow.
Notably, linearized power flow models have been advocated due to their simplicity, such as the well-known DC model~\cite{stott2009dc}, or the first-order approximation at an operating point for better accuracy~\cite{coffrin2014linear}.
As linear models are very limited by their generalizability across all possible operation region~\cite{trodden2013optimization},
one straightforward extension is the piecewise linear (PWL) approximation approach by using multiple operating points; see e.g.,~\cite{brown2020transmission}. By and large, there is a trade-off between accuracy and complexity for these model-based approaches, while the number and location of operating points could be difficult to select. 


To tackle these issues, data-driven approaches have been recently advocated as an alternative  for PWL modeling. Trained from realistic power flow scenarios, machine learning models such as $K$-plane regression~\cite{chen2021data} and neural networks (NNs) \cite{kody2022modeling,chen2023efficient} have shown good power flow approximation capabilities.
In particular, ReLU-based NNs can be used to construct simple yet accurate PWL power flow models by incorporating the power flow Jacobian information \cite{kody2022modeling}. Similar ideas have been explored in the constraint learning framework \cite{chen2023efficient} but for general grid operational constraints. Interestingly, these PWL models under the ReLU activation allow to reformulate grid optimization problems with nonlinear power flow as mixed-integer linear programs (MILPs), for which there exist efficient off-the-shelf solvers.  For example, successful applications to  unit commitment and distribution management have been considered. 
Albeit the success, these existing approaches mainly build on an end-to-end learning framework that does not consider the underlying physical models of power flow. In addition, none of them has yet considered a flexible grid topology.


In this paper, we put forth  a generative modeling approach to obtain the PWL approximation of AC power flow that is directly applicable to complex grid optimization problems.
With the ReLU activation, we  design the NN architecture to uniquely explore the generative structure of AC power flow.  
With the voltage and angle inputs, our proposed NN model first predicts the nonlinear terms that are common to all power variables in the first two layers, and then transforms these common terms to all line flows and power injections with two more layers. All the layers will be jointly trained to ensure an excellent consistency among all variables. This way, the proposed PWL model is generated in accordance with the power flow physics, while able to incorporate flexible topology connectivity.
Thanks to our proposed NN design,  we can cast grid optimization tasks like OTS 
into an MILP form for efficient solutions.
We use the IEEE 14- and 118-bus test systems to validate the proposed PWL approximation in terms of improving power flow modeling accuracy over non-generative, as well as an excellent optimality/feasibility performance in solving AC-OTS. 

The rest of the paper is organized as follows. Section~\ref{sec:lin} provides the nonlinear AC power flow modeling.
In Section~\ref{sec:piecewise}, we develop the PWL model that uses a neural network and discuss the formulation steps to attain a mixed-integer program.
Section~\ref{sec:sim} provides the simulation set-up for the IEEE 14- and 118-bus test systems and presents the numerical comparisons and validations for the proposed scheme, along with some concluding remarks.

\section{Nonlinear AC Power Flow Modeling} \label{sec:lin}

We first present the nonlinear AC power flow modeling for transmission systems, while introducing the relevant topology status variables and coupling terms useful for the discussions later on.
Consider a transmission system consisting of $n$ buses collected in the set $\mathcal{N} := \{1,\dots, n\}$ and $\ell$ lines (including transformers) in $\mathcal{L} := \{(i, j)\} \subset \mathcal{N} \times \mathcal{N}$. For each bus $i\in\mathcal{N}$, let $V_i\angle\theta_i$ denote the complex nodal voltage phasor, and  $\{P_i,Q_i\}$ denote the active and reactive power injections, respectively. For each line $(i,j)\in\mathcal{L}$, let $\theta_{ij}:= \theta_i-\theta_j$ denote the angle difference between bus $i$ and $j$, and $\{P_{ij},Q_{ij}\}$ denote the active and reactive power flows from bus $i$ to $j$; and similarly for $\{P_{ji},Q_{ji}\}$ from bus $j$ to $i$.
In addition, the line's series and shunt admittance values are respectively denoted by $y_{ij}= g_{ij}+\upj b_{ij}$ and $y_{ij}^{sh}=g^{sh}_i+\upj b^{sh}_i$.

By defining a binary variable $\epsilon_{ij} \in \{0,1\}$ to indicate the status for each line $(i,j)\in\mathcal{L}$ (0/1: off/on),  the nodal power balance at bus $i$  per the Kirchhoff's law becomes
\begin{subequations}\label{powerinj}
\begin{align}
P_{i} &= \textstyle \sum_{(i,j) \in \mathcal{L} } ~\epsilon_{ij} P_{ij},\\
Q_{i} &= \textstyle \sum_{(i,j) \in \mathcal{L} } ~\epsilon_{ij} Q_{ij}.
\end{align}
\end{subequations}
Note that these binary variables $\{\epsilon_{ij}\}$ will be important for formulating topology-related grid optimization tasks such as the optimal transmission switching problem as detailed later on. Without any topology changes, they can be fixed at $\epsilon_{ij}=1$. 
For each line $(i,j)\in\mathcal{L}$, the power flows relate to the angle difference $\theta_{ij}$ and nodal voltages $\{V_i, V_j\}$, and in the case of transformer, its tap ratio $a_{ij}$, as given by
\begin{subequations}
\label{powerflow}
\begin{align}
    P_{ij}&= V^2_{i} (\frac{ g_{ij}}{ a^2_{ij}}+ g^{sh}_i) - \frac{ V_{i}  V_{j}}{ a_{ij}} ( g_{ij}\cos  \theta_{ij}+ b_{ij}\sin  \theta_{ij}), \label{powerflow_1}\\
    Q_{ij}&= \! - V^2_{i} (\frac{ b_{ij}}{ a^2_{ij}}+ b^{sh}_i) \! -\! \frac{ V_{i}  V_{j}}{ a_{ij}} ( g_{ij}\sin  \theta_{ij} \!- \!b_{ij}\cos  \theta_{ij}).
\end{align}
\end{subequations}
For the transmission lines, we can simply set  $a_{ij}=1$. As for a transformer, the tap ratio is typically set within the range of $[0.9, 1.1]$ and it only affects the primary-to-secondary direction.  Thus, for the power flows in the secondary-to-primary direction, one can use $a_{ij}=1$ in \eqref{powerflow}.

One advantage of our proposed piecewise linear (PWL) approximation is to leverage the underlying coupling among active and reactive power flows. To this end, let us denote the three nonlinear terms in  \eqref{powerflow} by  
\[\gamma_i := V^2_i,~\rho_{ij}:= V_i  V_j \cos  \theta_{ij},~\mathrm{and}~\pi_{ij}:= V_i  V_j \sin  \theta_{ij}.\]
To form the bi-directional power flows $\{P_{ij},Q_{ij},P_{ji},Q_{ji}\}$ per line $(i,j)$,  we only need  the nonlinear terms $\{\gamma_i,\gamma_j,\rho_{ij},\pi_{ij}\}$, and the mapping between the two groups of variables is simply linear. 
To represent the resultant linear relation of \eqref{powerflow} in a matrix-vector form,  let us concatenate all the power flow variables in $\bm z^{pf} \in \mathbb{R}^{4\ell}$, and all the injection ones in $\bm z^{inj}\in \mathbb{R}^{2n}$. In addition, let $\bm \gamma \in \mathbb{R}^{n}$, $\bm \rho \in \mathbb{R}^{\ell}$, and $\bm \pi \in \mathbb{R}^{\ell}$ denote the respective vectors for the three groups of common terms. This way, we have 
\begin{subequations}\label{pow}
\begin{align}
    \bm z^{pf} &= \bm W^{\gamma} \bm \gamma + \bm W^{\rho} \bm \rho + \bm W^{\pi} \bm \pi,\label{powa}\\
    \bm z^{inj} &= \bm W^{\psi} \bm z^{pf}
\label{powb}
\end{align}
\end{subequations}
where the weight matrices $\{\bm W^{\gamma}, \bm W^{\rho}, \bm W^{\pi}, \bm W^{\psi}\}$ are of appropriate dimension  given by the known line parameters and line status variables in \eqref{powerflow} and \eqref{powerinj}, respectively.
Clearly, the three groups of  common terms are sufficient for fully  generating  all power flow and injection quantities, and our proposed generative modeling will work by predicting these terms as the first step.

\subsection{Linear Approximation}
\label{sec:linappx}
We discuss the linear approximation for the common terms, which will be used by the proposed PWL models. 
Linear approximation is a basic approach to deal with power flow nonlinearity, thanks to its simplicity and reasonable accuracy within a small region of the operating point. We will consider the first-order approximation method to attain a linearized modeling, while there also exist other popular methods such as fixed-point method~\cite{simpson2017theory}. 

For the squared voltage term, it can be approximated by  $\hat \gamma_i=  2V_i - 1$, $\forall i\in\mathcal{N}$, based on a flat-voltage value of $V_o$. Of course, this linearized model can be improved by using the exact operating point if different from the flat-voltage profile. As shown in~\cite{trodden2013optimization}, the former already attains a very high accuracy for power systems with well-regulated bus voltages within the p.u. range of [0.94, 1.06]. Thus, for simplicity, this linear representation of $\hat {\bm{\gamma}}$ will be adopted in this work.

Nonetheless, the other two terms $\bm \rho$ and $\bm \pi$ are more complicated to approximate than $\bm \gamma$ due to the presence of angle differences. To this end, we consider the first-order approximation for the former at the operating point. 
To simplify the notation, 
let us use $[\bm \rho;~\bm \pi] = \bm f(\bm x) \in \mathbb R^{2\ell}$ to represent the nonlinear mapping from the input $\bm x$, which consists of the voltage magnitude $\bm V =\{V_i\}_{i\in\mathcal N} \in \mathbb R^{n}$ and the angle difference  $\bm \theta =\{\theta_{ij}\}_{(i,j)\in\mathcal{L}} \in \mathbb R^{\ell}$.
With a fixed operating point denoted by $\bm x_o$,  the first-order approximation becomes  
\begin{align}
    [\hat{\bm \rho}; \hat{\bm \pi}]&= \bm f(\bm x_o)+ \bm J(\bm x_o)(\bm x - \bm x_o) \nonumber\\
    &= \bm f(\bm x_o)+ \bm J(\bm x_o)\Delta \bm x \label{linearf}
\end{align}
where $\bm J(\bm x_o)$ denotes the Jacobian matrix of $\bm f(\bm x)$ evaluated at  $\bm x_o$, while we use $\Delta \bm x := \bm x -\bm x_o$ for simplicity. 
The ensuing section will build upon the linear model in \eqref{linearf} by using a data-driven approach to improve the approximation accuracy.  

\begin{figure}[t!]
	\begin{center}
		\includegraphics[scale=0.33]{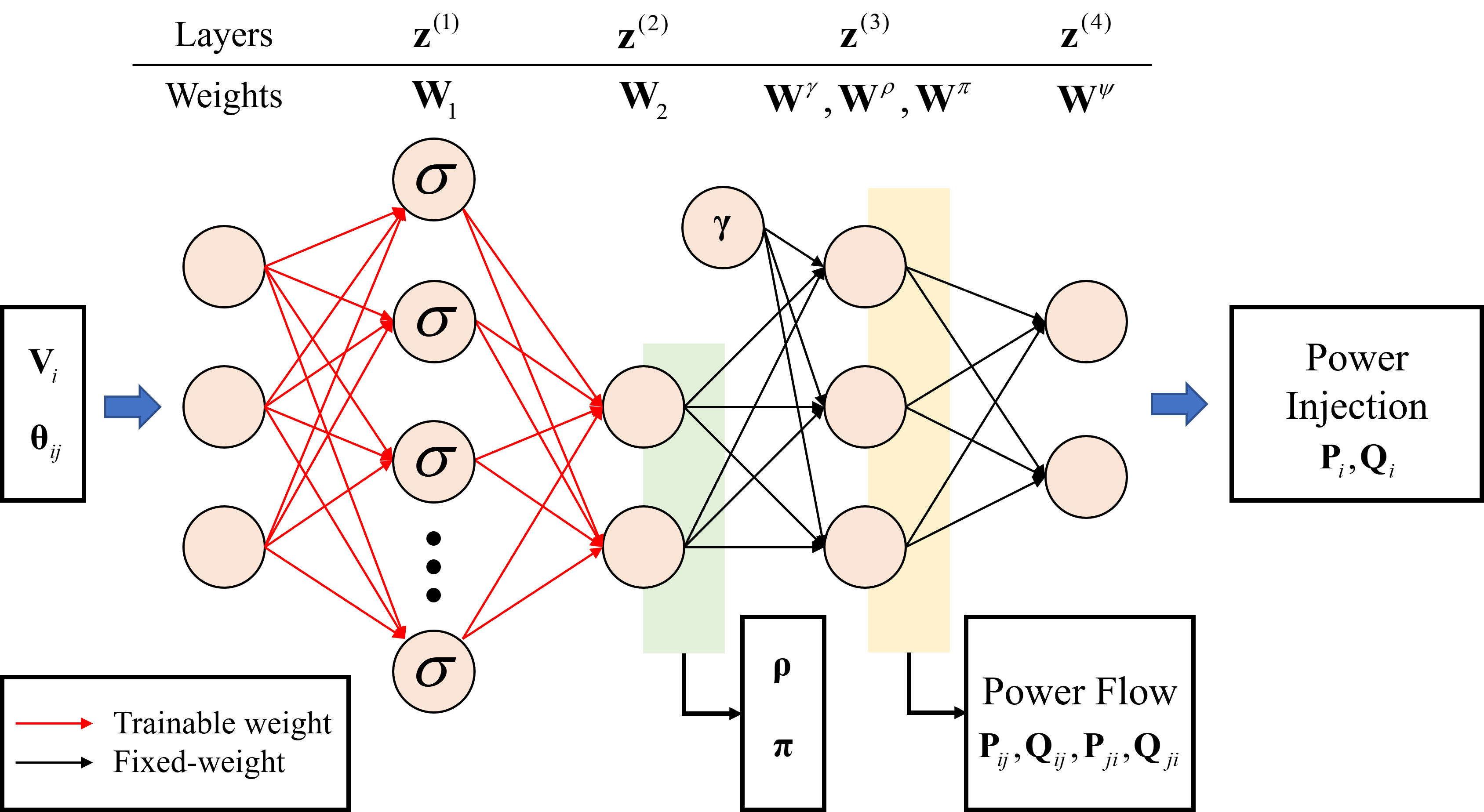}
		\caption{The structure of the proposed neural network that can generate the common terms on the second layer through the trainable weight matrix and generate power flows and injections on the third and fourth layers through the fixed weight matrices.} \label{structure}
	\end{center}
\end{figure}

\section{PWL Approximation via Generative Modeling} \label{sec:piecewise}

Our proposed PWL model uses a two-layer neural network (NN) to first approximate the common nonlinear terms in  $[\bm \rho;\bm \pi]$, followed by two additional linear layers to generate the power flow and injection variables. 
As illustrated in Fig.~\ref{structure}, using the input voltage and angle difference in $\bm x =[\bm V;\bm\theta]$, the first two layers will rely on the ReLU activation functions to form the best PWL model for $[\hat{\bm \rho}; \hat{\bm \pi}]$  by adjusting the NN parameters. In addition, the last two layers use the fixed weight parameters from \eqref{powa} to generate the power flows for all lines, and accordingly, use \eqref{powb} to generate the power injection at all nodes. Note that the squared voltage $\hat{\bm \gamma}$ used by the third layer of generating power flow is based on the simple linearized model as described in Sec.~\ref{sec:linappx}. Thus, the proposed approximation fully matches the power flow relations and coupling among different terms, in an efficient and generative fashion. 

Using the ReLU activation, the first two layers can effectively produce a PWL mapping that can improve the accuracy of the linearized model in \eqref{linearf}. The ReLU function is defined by $\sigma(\cdot)$ where outputs the entry-wise maximum between the input value and 0.
Intuitively, when the activation status of ReLU function stays unchanged within a certain region of input values, its functional output enjoys the same linear relation with the input in that region.
Therefore, one can view that the combination of ReLU activation status would divide the whole input space into multiple smaller regions, within each it boils down to a purely linear function. Thus, the overall function over the whole input space becomes a PWL one. In this sense, the number of linear regions will grow exponentially with the number of activation functions, and it would be challenging to search for all possible combinations. Therefore, we will train the NN parameters within the first two layers from generated data samples that can best select the activation status and linear regions from the data. 

To concretely connect the NN model with PWL functions, we first consider a simple case of two linear regions obtained by using two different operating points, namely $\bm x_o$ and $\bm x_1$, as
\begin{align}\nonumber
    [\hat{\bm \rho}; \hat{\bm \pi}]& =   \bm f(\bm x_o) + \Delta \bm y,~\mathrm{with}\\
    \label{dy} \Delta \bm y &= \begin{cases}
    \bm J(\bm x_o)\Delta\bm x, & \bm x \in \mathcal{R}_o \\[1\jot]
    \bm J(\bm x_1)(\bm x - \bm x_1) + \bm r, & \bm x \in \mathcal{R}_1
    \end{cases} 
\end{align}
where $\mathcal{R}_q$ represents the linear region corresponding to $\bm x_q$, and  the residue in $\mathcal{R}_1$ is given by $ \bm r := \bm f(\bm x_1) - \bm f(\bm x_o)$.

To recover the NN structure for \eqref{dy}, we follow \cite{eckart1936approximation} to assume that the two Jacobian matrices therein are different by a low-rank component. Specifically, we assume that 
\begin{align}
    \bm J(\bm x_1) \approxeq \bm J(\bm x_o)+ \bm w_2 \bm w^\top_1 \label{Jac2}
\end{align}
where both $\bm w_1$ and $\bm w_2$ consist of the NN weight parameters, that can be of much lower dimension than the size of Jacobian matrix.
For simplicity, we consider both of them to be vectors with $\bm w_1 \in \mathbb R^{n+\ell}$ and $\bm w_2 \in \mathbb R^{2\ell}$, and thus the difference term in \eqref{Jac2} becomes a rank-one matrix. This will be expanded to a higher-rank case later by using weight matrices. 
Interestingly, the simplification in \eqref{Jac2} allows to unify the two scenarios in \eqref{dy} by using one ReLU activation function, as given by 
\begin{align}
    \Delta \bm y \approxeq \bm J(\bm x_o)\Delta \bm x + \bm w_2 \sigma(\bm w^\top_1 \Delta \bm x+ b)\label{one_q}
\end{align}
where $b$ is a scalar bias parameter.
When the ReLU function is not activated, it becomes the linear model in $\mathcal R_o$ of \eqref{dy}. Otherwise, upon the activation of $\sigma(\cdot)$ the resultant linear model should approach the one in $\mathcal R_1$ by recognizing the relation between the two Jacobian matrices in \eqref{Jac2}, as given by
\begin{align}
    \Delta \bm y \approxeq \Big(\bm J(\bm x_o)+ \bm w_2 \bm w^\top_1 \Big)\Delta \bm x + \bm w_2 b.
\end{align}
In addition, to match the offset term in $\mathcal R_1$,  we would need to have $\bm w_2 b \approxeq \bm J(\bm x_1)(\bm x_o - \bm x_1) + \bm r$.
In general, the two-layer form in \eqref{one_q} may not fully express or match the two-region linearized model at the two operating points as in \eqref{dy}. Nonetheless, \eqref{one_q} definitely constitutes as a PWL approximation for the underlying $\bm f(\bm x)$ function. In particular, the single ReLU activation in \eqref{one_q} has led to 2 linear regions for the resultant PWL model. 

The simple case of two linear regions can be expanded to encompass more complex PWL model by increasing the number of ReLU activation functions. If the first layer has $q$ ReLU functions with different linear transformations as the input, it is possible  to generate a PWL model with up to $2^q$ linear regions. 
This way, the weight parameters form the two matrices $\bm W_1 \in \mathbb R^{(n+\ell)\times q}$ and $\bm W_2 \in \mathbb R^{2\ell\times q}$, as well as the bias vector $\bm b \in \mathbb R^{q}$. The number of linear regions is related to the combination of activation status for all $q$ ReLU functions. 
The larger $q$ is, the more expressive the corresponding PWL model becomes, at the price of more model parameters to consider. This makes it difficult to determine the weight parameters using model-based linearization as in \eqref{dy}, motivating us to train these parameters from generated power flow samples. In general, the latter can be designed to reflect the realistic operating points and the statistical variability around them, and thus the resultant PWL model could outperform a model-based approach by pre-selecting the points for linearization.

Before presenting the training loss, recall that the full generative model in Fig.~\ref{structure} includes the first two layers for obtaining the nonlinear terms and two fixed-weight layers for power variables, as given by 
\begin{subequations}\label{NN_pf}
\begin{align}
    \bm z^{(1)} &= \sigma(\bm W^\top_1 \Delta \bm x + \bm b),\label{NN_pf1}\\
    \bm z^{(2)} &= \bm f(\bm x_o) + \Big( \bm J(\bm x_o)\Delta \bm x +  \bm W_2 \bm z^{(1)} \Big),\label{NN_pf2}\\
    \bm z^{(3)} &= \bm W^{\gamma} \hat{\bm \gamma} + [\bm W^{\rho};\bm W^{\pi}]\bm z^{(2)},\label{NN_pf3}\\
    \bm z^{(4)} &= \bm W^{\psi} \bm z^{(3)}\label{NN_pf4}
\end{align}
\end{subequations}
where the first two layers generalize the simple case of \eqref{one_q} to $q$ ReLU functions, and the last two layers follow from \eqref{pow}. When we generate random power flow data, the actual values for both $\bm f(\bm x)= [\bm \rho; \bm \pi]$ and $[\bm z^{pf}; \bm z^{inj}]$ can be obtained and using all of them for the loss function could effectively maintain the relations among the corresponding predicted values in \eqref{NN_pf}. Specifically, we can use the Euclidean distance to form the following loss function 
\begin{align}
    \mathcal{L}({\bm W_1, \bm W_2, \bm b})= &\| \bm f(\bm x) - \bm z^{(2)}  \|^2_2 \nonumber \\
    &+\lambda \left\| [\bm z^{pf}; \bm z^{inj}] -  [\bm z^{(3)}; \bm z^{(4)}] \right\|^2_2 \label{update}
\end{align}
where $\lambda>0$ denotes a regularization hyperparameter to balance the error terms among the layers. The average loss will be used to aggregate different samples, yielding the total training loss objective to minimize.
After training, the proposed PWL model can fully generate the power flows and injections with linear transformations.
On the other hand, \eqref{NN_pf1} is not a linear transformation due to the ReLU function.
We also face nonlinear constraints when considering the binary line status variables with power flows.
Hence, we will work to reformulate the ReLU function and the line status-related constraints into mixed-integer linear forms.

\subsection{Mixed-integer Linear Formulation for the PWL Model}

The proposed PWL models allow for formulating the nonlinear power flow equations into mixed-integer linear forms and thus enable efficient solutions for grid optimization problems  involving topology variables. We will present the formulation steps to attain a mixed-integer linear program (MILP) for the PWL models, and also for dealing with the binary line status relation in \eqref{powerinj}.

To adopt the PWL model in \eqref{NN_pf} into an MILP, the main issue lies in the ReLU function of \eqref{NN_pf1}, as all other transformations are just linear ones.  
To tackle the ReLU function, we will use a relaxation technique based on the big-M tightening method~\cite{griva2009linear}.
For the $k$-th entry $z^{(1)}_k$ in \eqref{NN_pf1}, we will approximate it by introducing a binary variable $\beta_k$, and its upper/lower bounds $\{\bar M_k,\underline M_k\}$. 
The two bounds can be determined through an off-line optimization procedure~\cite{grimstad2019relu}. After determining these bounds and denoting the input in \eqref{NN_pf1} by $\hat {\bm z}^{(1)} = \bm W^\top_1 \Delta \bm x + \bm b$, the big-M method asserts that each $z^{(1)}_k$ can be reformulated by using four linear inequality constraints, as given by
\begin{subequations}\label{MILP}
\begin{align}
    0 &\leq  z^{(1)}_k \leq \bar M_k  \beta_k, \label{beta0}\\
    \hat {z}^{(1)}_k &\leq z^{(1)}_k \leq \hat {z}^{(1)}_k - \underline M_k (1- \beta_k), \label{beta1}
\end{align}
\end{subequations}
The binary variable $\beta_k$ critically relates to the ReLU activation status based on the input $ \hat{z}^{(1)}_k$.
If the input $\hat {z}^{(1)}_k > 0$, then  the constraints in  \eqref{beta1} enforce $\beta_k$  to be one such that $z^{(1)}_k=\hat {z}^{(1)}_k$ holds exactly.
Otherwise, if $\hat {z}^{(1)}_k \leq 0$, the constraints in \eqref{beta0} enforce $\beta_k$ to be zero to yield $z^{(1)}_k=0$.
This way, the output $z^{(1)}_k$ from \eqref{MILP} exactly attains the ReLU-based output in \eqref{NN_pf1}. Thus, with accurate upper/lower bounds,  the big-M method allows for an equivalent reformulation  of \eqref{NN_pf} into an MILP form.

Similarly, we formulate the line status-related constraints into an MILP form.
We face the multiplication of continuous variables $\{P_{ij}$, $ Q_{ij}\}$ and binary variables $ \epsilon_{ij}$ that are denoted as $\hat{ P}_{ij} := \epsilon_{ij}  P_{ij}$ and $\hat{ Q}_{ij} := \epsilon_{ij}  Q_{ij}$.
To tackle this multiplication term, we will use the McCormick relaxation technique~\cite{mccormick1976computability} derived from the big-M tightening method.
After attaining the upper/lower bounds of active and reactive power flows $\{{\bar{ P}}_{ij},{\underline{ P}}_{ij}\}$ and $\{{\bar{ Q}}_{ij},{\underline{ Q}}_{ij}\}$, each $\{\hat{ P}_{ij},\hat{ Q}_{ij}\}$ can be reformulated by using four linear inequality constraints, as given by
\begin{subequations}\label{sw}
\begin{align}
    {\underline{ P}}_{ij}  \epsilon_{ij} &\leq \hat{ P}_{ij} \leq {\bar{ P}}_{ij}  \epsilon_{ij}, \label{sw0_P}\\
    {\underline{ Q}}_{ij}  \epsilon_{ij} &\leq \hat{ Q}_{ij} \leq {\bar{ Q}}_{ij}  \epsilon_{ij}, \label{sw0_Q}\\
     P_{ij} + {\bar{ P}}_{ij} ( \epsilon_{ij}-1) &\leq \hat{ P}_{ij} \leq  P_{ij} + {\underline{ P}}_{ij} ( \epsilon_{ij}-1), \label{sw1_P}\\
     Q_{ij} + {\bar{ Q}}_{ij} ( \epsilon_{ij}-1) &\leq \hat{ Q}_{ij} \leq  Q_{ij} + {\underline{ Q}}_{ij} ( \epsilon_{ij}-1). \label{sw1_Q}
\end{align}
\end{subequations}
The outputs $\{\hat{ P}_{ij},\hat{ Q}_{ij}\}$ critically relate to $\epsilon_{ij}$.
If $ \epsilon_{ij}$ is equal to zero, then  the constraints in \eqref{sw0_P} and \eqref{sw0_Q} enforce $\hat{ P}_{ij}$ and $\hat{ Q}_{ij}$ to be zero.
Otherwise, if $ \epsilon_{ij}$ is equal to one, then  the constraints in \eqref{sw1_P} and \eqref{sw1_Q} enforce $\hat{ P}_{ij}$ and $\hat{ Q}_{ij}$ to be $ P_{ij}$ and $ Q_{ij}$, respectively.
This way, the output $\hat{ P}_{ij}$ and $\hat{ Q}_{ij}$ from \eqref{sw} exactly attain the power flows based on the binary line status.
Thus, the McCormick relaxation allows for an equivalent reformulation of the multiplication terms of power flows and line status variables into an MILP form.

\section{Numerical Studies} \label{sec:sim}

We have implemented the proposed generative modeling approach on the IEEE 14-bus and 118-bus test cases~\cite{IEEE_case_ref}, to compare its performance in power flow modeling and grid topology optimization.
The NN training has been performed in PyTorch with Adam optimizer on a regular laptop with Intel\textsuperscript{\textregistered} CPU @ 2.70 GHz, 32 GB RAM, and NVIDIA\textsuperscript{\textregistered} RTX 3070 Ti GPU @ 8GB VRAM.
We have formulated the OTS problem through Pyomo~\cite{hart2011pyomo} and used the Groubi optimization solver~\cite{gurobi} for the resultant MILPs.

To train the proposed NN-based PWL models in Fig.~\ref{structure}, 
we generate 10,000 samples from the actual power flow model, with the outputs of common nonlinear terms $\{\bm \gamma, \bm \rho, \bm \pi\}$, as well as line flows and nodal injections.
For each sample, we generate uniformly distributed voltage magnitudes within the range of [0.94,~1.06] p.u., 
and similarly for the angle, which randomly varies within $[-\pi/6, \pi/6]$ radians around the initial operating point.
For the reference bus, namely \texttt{Bus 1} in the 14-bus or \texttt{Bus 69} in the 118-bus system, we fix its voltage magnitude and angle at default values.
For the first two trainable layers in Fig.~\ref{structure}, we use $q=25$ and $q=75$ ReLU activation functions, respectively for the two systems. 
The parameters are trained through the backpropagation using the loss function in \eqref{update} with $20\times10^3$ epochs and a learning rate of $2.5\times10^{-3}$.
For the training of the PWL model, we separate 90\% of the data set as training and 10\% as testing.
The PWL model only trains the training dataset and validates the loss through the testing dataset.

\begin{figure}[t!]
	\centering
	\subfloat[Average Error]{\includegraphics[scale=0.25]{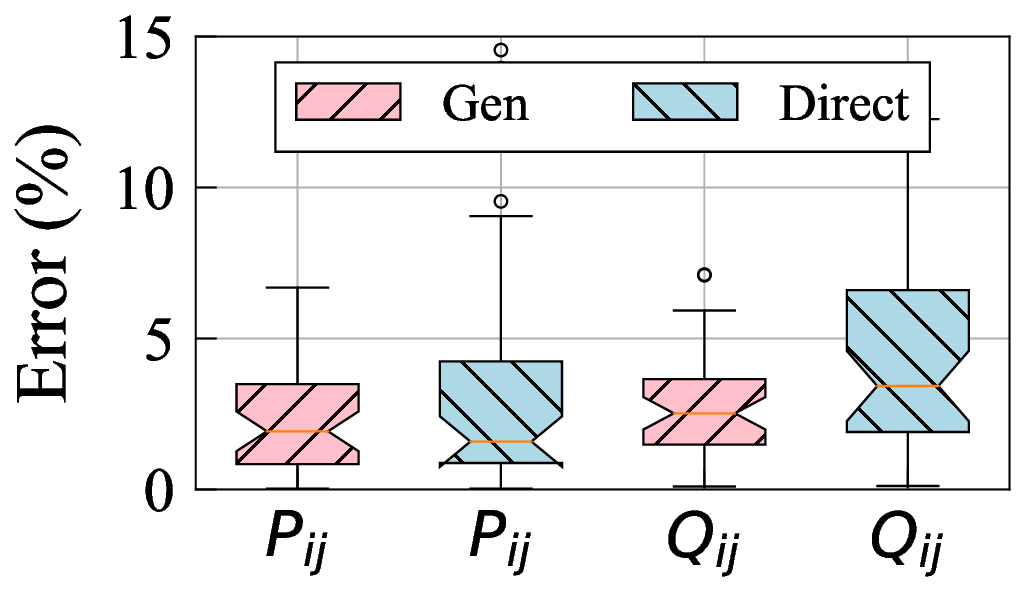}\label{avg_14}}
	\quad
	\subfloat[Maximum Error]{\includegraphics[scale=0.25]{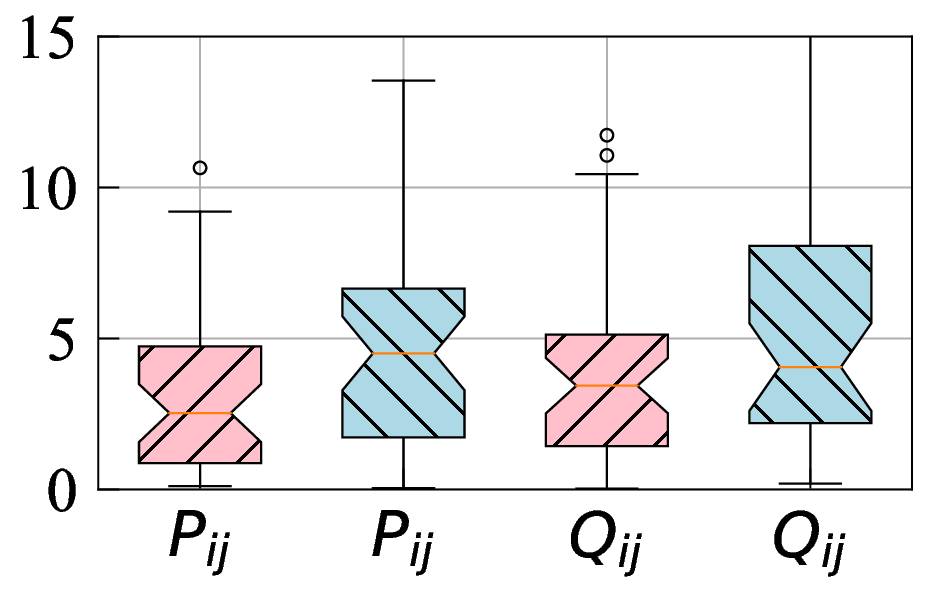}\label{max_14}}
	\caption[]{\small Comparisons of the (a) average error and (b) maximum error in approximating the line power flows  for both the proposed Gen and Direct methods  using the IEEE 14-bus system.}\label{ACPF14}
\end{figure}

\begin{figure}[t!]
	\centering
	\subfloat[Average Error]{\includegraphics[scale=0.25]{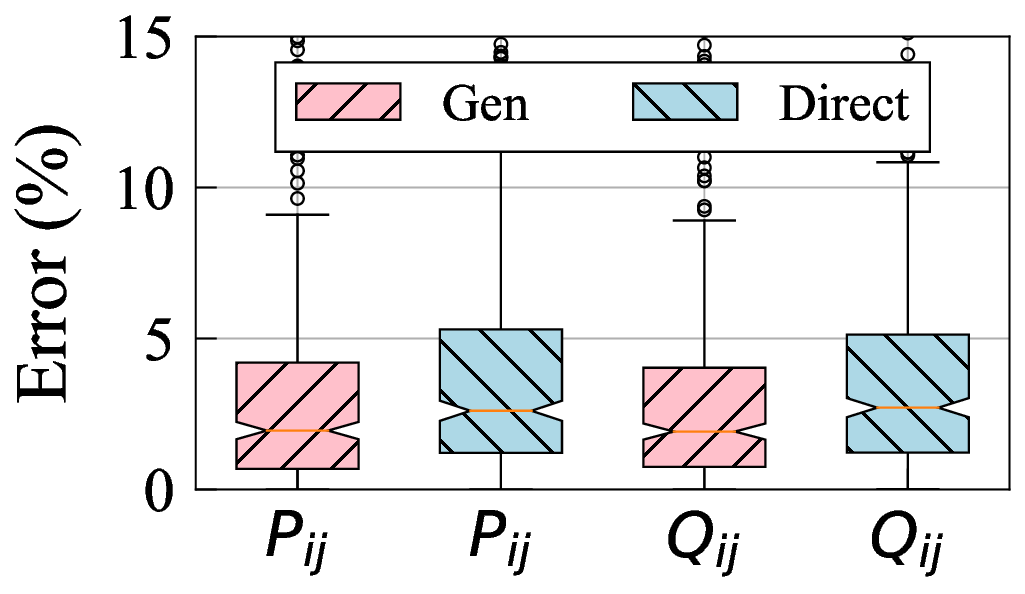}\label{avg_118}}
	\quad
	\subfloat[Maximum Error]{\includegraphics[scale=0.25]{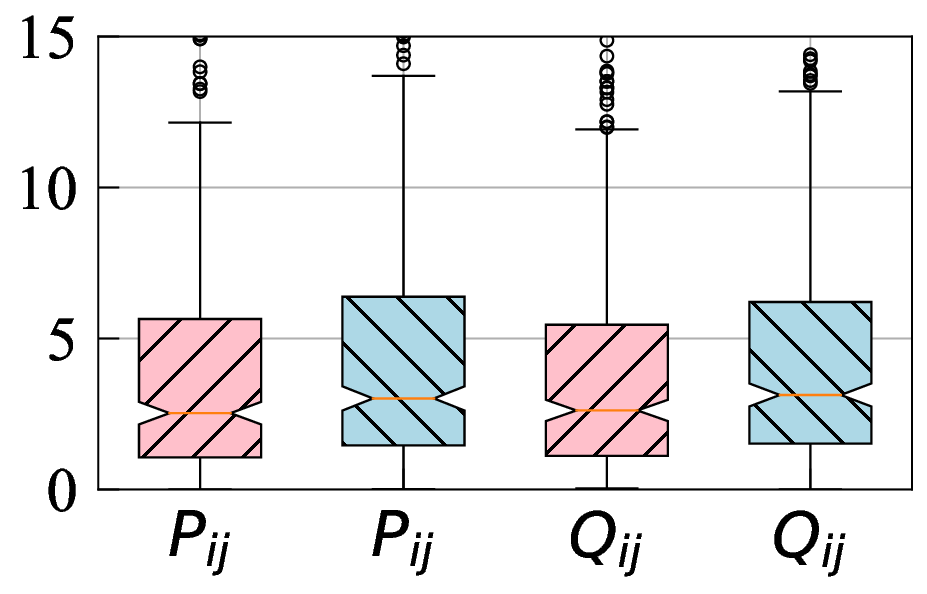}\label{max_118}}
	\caption[]{\small Comparisons of the (a) average error and (b) maximum error in approximating the line power flows  for both the proposed Gen and Direct methods  using the IEEE 118-bus system.}\label{ACPF118}
\end{figure}

\subsection{AC Power Flow Approximation}
We first validate the AC power flow modeling performance of our PWL-based approximation. We compare the proposed generative modeling approach using the common nonlinear term prediction step (indicated by Gen) with the existing work \cite{kody2022modeling} that directly predicts the  power flow variables (indicated by Direct). The latter directly uses a two-layer NN of ReLU activation to output the line power flow, with the structure given by 
\begin{align}\label{dNN_pf}
    \bm z ^{pf}&\leftarrow \bm J(\bm x_o)\Delta \bm x +  \bm W_2 \bm z^{(1)}.
\end{align}
Note that we use the same number of activation functions for both types of models.

We compare the approximation error between the predicted and actual line power flow values as  normalized by the line capacity.
Figs.~\ref{ACPF14} and~\ref{ACPF118} show the box plots of the normalized prediction error percentages of both active and reactive line flows, respectively for the  14- and 118-bus systems. Note that both the average error and the maximum error, out of all transmission lines in each system are included for comparisons. 
Each box plot shows the median values as midlines, the first and third quartiles as boxes, maximum values as horizon bars, and some outliers.  
Clearly, the proposed generative model is of better accuracy in predicting the line flows than the direct method, especially for the reactive power parts.
Notably, the proposed method has shown significant improvements in terms of reducing the maximum values of errors in all cases.
These results have verified the benefits of incorporating the underlying coupling between active and reactive power flows considered by our proposed NN design. 

\begin{figure}[t!]
	\centering
	\subfloat[IEEE 14 bus]{\includegraphics[scale=0.25]{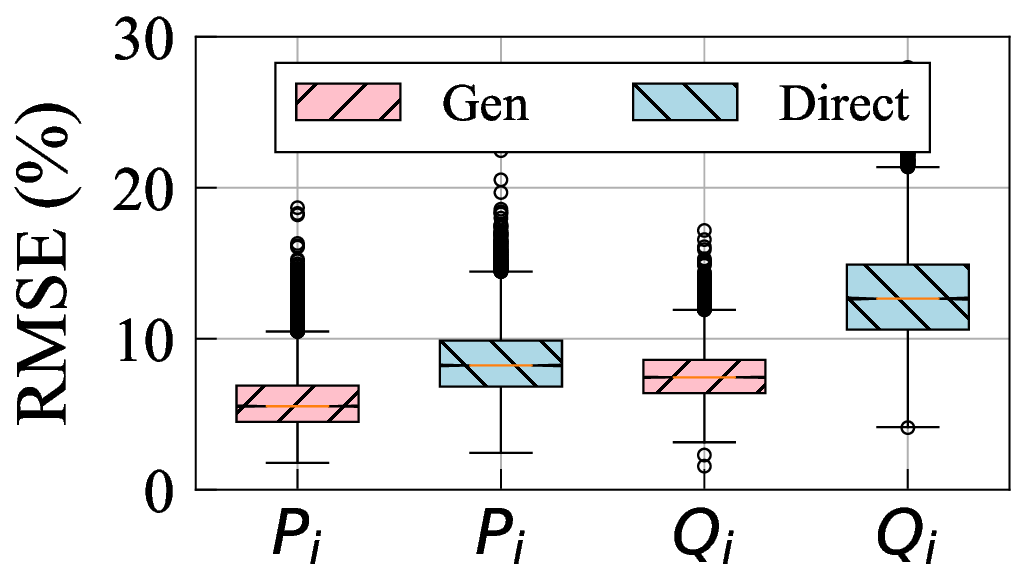}\label{inj_14}}
    \quad
	\subfloat[IEEE 118 bus]{\includegraphics[scale=0.25]{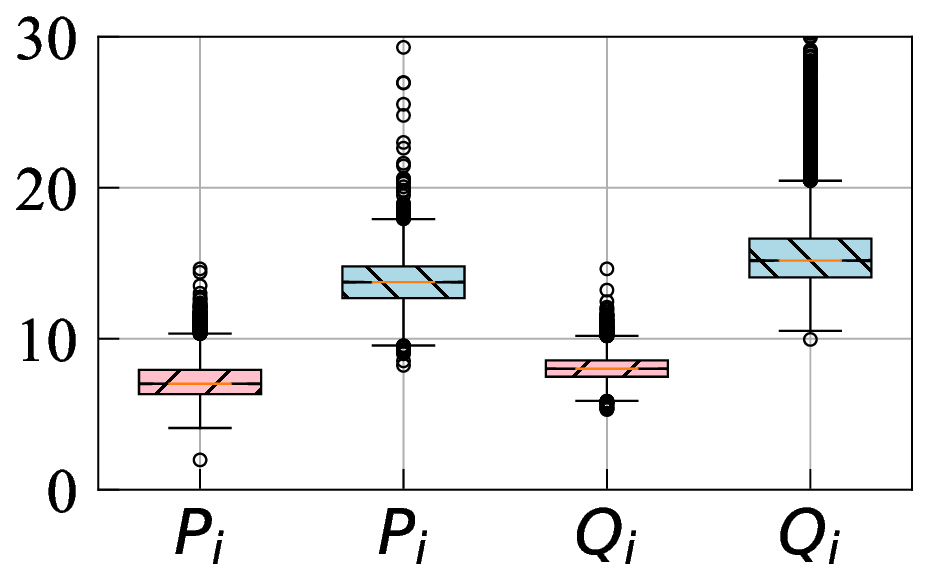}\label{inj_118}}
	\caption[]{\small Comparisons of the root mean square error (RMSE) in predicting the injected power vectors for both the proposed Gen and Direct methods  in the (a) IEEE 14-bus and (b) 118-bus systems.}\label{ACPFinj}
\end{figure}

Furthermore, we also compare the error performance in predicting the active and reactive power injections, which can be formed directly from the line flows using \eqref{powb}.
Without any normalization basis, Fig.~\ref{ACPFinj} instead shows the box plots for the root mean square error (RMSE) in predicting the injected power vectors in both test systems. 
Similar results have been observed for predicting the injections, with even more noticeable improvements in both active and reactive power values. This is because our proposed NN model in Fig.~\ref{structure} has directly accounted for the power flow coupling, and thus its joint training process would achieve high consistency with nodal power balance. Thanks to the generative structure of the underlying NN design, our proposed PWL models can improve the accuracy and consistency in the resultant power flow approximation. 


\subsection{OTS Applications}
We adopt the proposed PWL models in solving the OTS problem using the 118-bus system.
The objective function and operational constraints are set up similarly  to the typical optimal power flow (OPF) problem. Additionally, OTS allows for line switching under the constraint of a total switching budget of $\alpha$ lines, given by
\[\textstyle \sum_{(i,j) \in \mathcal{L} } ~\epsilon_{ij} \geq \ell-\alpha.\]
The number $\alpha$ is typically no greater than 5-10. 
Hence, the computation complexity of OTS is much higher than that of OPF, due to the integer line status variables. 
We introduce the proposed PWL models into the AC-OTS formulation by replacing the power flow constraints by \eqref{MILP}, as well as the line status constraints by \eqref{sw}.
This way, the resultant MILP problem can be efficiently solved  with solvers like Gurobi.

We test the performance of the proposed PWL model-based OTS solutions.
We compare it with the OTS solutions using the DC- and AC- power flow models, both provided by the open-source platform~\cite{8442948}. To compare across different OTS methods, we re-run the AC-OPF problem after fixing the topology with their line-switching decision outputs, using the MATPOWER~\cite{zimmerman2010matpower} solver. This way, we can compare the metrics in terms of the objective costs (for optimality), as well as the percentage rates of infeasibility and constraint violations (for feasibility), using the corresponding AC-OPF outputs. For the two feasibility measures, the infeasible solution rates measure the percentage of infeasible solutions over all solutions, while the constraint violation rates are based on  the percentage of over-limit voltage magnitude and angle over the infeasible solutions.
Table~\ref{PLOTS} lists these metrics and also the computation time for each of the three OTS methods with a switching budget $\alpha$ equal to 1 or 3. Note that the computation time corresponds to solving the OTS optimization problem, not the follow-up AC-OPF one. 
A total of 1,000 power flow scenarios by having nodal demand  uniformly distributed within [50\%, 200\%] of the initial demands~\cite{IEEE_case_ref} have been used to compute the average of all these four metrics. For the objective cost, the AC-OTS method has been used as a baseline (normalized to be 100\%), and thus the other two OTS methods using approximate models attain higher percentage values. Nonetheless, the proposed PWL model only slightly increases the objective cost by less than 2\%, while attaining exactly the same infeasiblity metrics as the AC-based OTS solutions. Notably, our model achieves a highly competitive performance and also great efficiency, as its computation time is almost a tenth of the AC-OTS one. In particular, the PWL model has allowed for a very low computation complexity in the order of DC-OTS one. But the latter leads to significantly worse feasibility performance, with an almost order of magnitude higher of infeasiblity rates than PWL-based OTS.  Thanks to its high modeling accuracy, our proposed PWL model can greatly simplify the computation for grid topology optimization tasks by using the MILP reformulation trick, while approaching the ideal optimality/performance performance.

\begin{table}[t!]
\caption{Comparison of the optimality and feasibility of the solutions of the AC, PWL, and DCOTS.}
\begin{center}
\begin{tabular}{c|c|c|c|c}
\Xhline{3\arrayrulewidth}
  &  {\makecell[c]{Switching\\Budget}}  & AC & PWL & DC \\ \hline
\multirow{2}{*}{\makecell[c]{Objective\\ Cost (\%)}} & $\alpha=1$ &   100\%       &    101.90\%    &   102.67\% \\ \cline{2-5}
 &  $\alpha=3$ &  100\%       &    101.94\%    &   104.21\% \\ \hline
\multirow{2}{*}{\makecell[c]{Infeasible\\Solution (\%)}} & $\alpha=1$ & 0.20\% & 0.20\% & 2.10\% \\ \cline{2-5}
 & $\alpha=3$ & 0.20\% & 0.20\% & 2.80\% \\ \hline
\multirow{2}{*}{\makecell[c]{Constraint\\Violation (\%)}} & $\alpha=1$ & 0.33\%     &  0.33\%   & 1.32\%   \\ \cline{2-5}
 & $\alpha=3$ & 0.33\%     &  0.33\%   & 2.31\%   \\ \hline
\multirow{2}{*}{\makecell[c]{Computation\\Time (s)}}  & $\alpha=1$ &    52.13 s    &  5.37 s  & 3.41 s \\ \cline{2-5}
  & $\alpha=3$ &    57.67 s    &  6.45 s  & 3.89 s \\
\Xhline{3\arrayrulewidth}
\end{tabular}\label{PLOTS}
\end{center}
\end{table}

To sum up, we have designed a NN-based PWL approximation model for AC power flow with a good balance between model complexity and accuracy. Through its generative design, the proposed PWL models not only account for the underlying power flow coupling, but also allow for highly competitive solutions for  topology-aware grid optimization problems. 
Our future research directions include improving the scalability of our proposed PWL models in large-scale power systems, as well as considering more generalized topology-aware grid optimization tasks like restoration and adaptive islanding.

\ifCLASSOPTIONcaptionsoff
\newpage
\fi
	
\bibliographystyle{IEEEtran}
\bibliography{Ref.bib}

\end{document}